\documentclass{aastex62}
\usepackage[nointegrals]{wasysym}
\usepackage{amsmath,upgreek}

\graphicspath{{./}{figures/}}

\submitjournal{and accepted by the Serbian Astronomical Journal}

\shorttitle{Dyson Spheres}
\shortauthors{Jason T.\ Wright}

\begin{document}

\title{Dyson Spheres}

\correspondingauthor{Jason Wright}
\email{astrowright@gmail.com}

\author[0000-0001-6160-5888]{Jason T.\ Wright}
\affil{Department of Astronomy \& Astrophysics and \\
Center for Exoplanets and Habitable Worlds\\ 
and Penn State Extraterrestrial Intelligence Center \\
525 Davey Laboratory \\
The Pennsylvania State University \\
University Park, PA, 16802, USA}



\begin{abstract}

I review the origins and development of the idea of Dyson spheres, their purpose, their engineering, and their detectability. I  explicate the ways in which the popular imagining of them as monolithic objects would make them dynamically unstable under gravity and radiation pressure, and mechanically unstable to buckling. I  develop a model for the radiative coupling between a star and large amounts of material orbiting it, and connect the observational features of a star plus Dyson sphere system to the gross radiative properties of the sphere itself. I  discuss the still-unexplored problem of the effects of radiative feedback on the central star's structure and luminosity. Finally, I  discuss the optimal sizes of Dyson spheres under various assumptions about their purpose as sources of low-entropy emission, dissipative work, or computation.

\end{abstract}



\section{Introduction}

\subsection{Dyson Spheres}

The physics and observational consequences of Dyson spheres were first described by their eponym \citet{dyson60} at the dawn of modern SETI. \citeauthor{dyson60}'s original suggestion was quite general, pointing out that the exponential growth of an extraterrestrial species' energy supply could lead it to occupy ``an artificial biosphere which completely surrounds its parent star.'' He argued that the waste heat of the ``energy metabolism'' of this extraterrestrial technology would likely be detectable at 10$\upmu$, and that a search of the sky at mid-infrared wavelengths for point sources of such emission would be worthwhile. 

In response to letters to the editor reacting to his work, \citet{dysonletters} clarified that 1) he had not envisioned a monolithic shell or ring---which he wrote would be ``mechanically impossible''---but rather a ``loose collection or swarm of objects traveling on independent orbits around the star;'' 2) that his description of infrared emission was independent of the specifics of how or why such a swarm might be constructed; and that 3) detection of infrared excess around nearby stars ``would not by itself imply that extraterrestrial intelligence had been found,'' but would be an important discovery regardless of its cause.

\citet{Dyson66} later elaborated on the engineering behind the idea, showing there is no physical barrier to building large rigid structures in space (where ``large'' was up to $10^6$ km), that such structures can be very light, and that planets can (in principle) be deconstructed for their material in sufficient qunatities to build a sphere of useful thickness (of order meters). He also outlined how such structures would be discovered: first by their 3-10$\upmu$ excess consistent with being a protostar,\footnote{``When such objects are found, I hope nobody will rush to the newspapers with claims that we have found something artificial. I say only that if highly expanded technologies exist, they are to be found among such objects\ldots Certainly one will not claim any object to be artificial unless it appears pathological in at least two independent modes of observation.'' \citep{Dyson66}} then diagnosed as artificial via ``irregular light variations due to starlight shining through chinks in the curtain, and for stray electromagnetic fields and radio noise produced by large-scale electrical operations.'' 

But unlike the radio program envisioned by \citet{cocconi_searching_1959} and executed by \citet{OZMA} in 1960, searches for technology along the lines of \citeauthor{dyson60}'s suggestion are still in their infancy, and indeed the observational consequences of such technology are still only roughly understood. \citet{Bradbury01} presents a good analysis of the history of the idea, the confusion in the literature regarding the shape and purpose of the technology in \citeauthor{dyson60}'s original suggestion, and an extensive catalog of the literature on the subject. In particular, he examines the assumptions that the sphere would necessarily constitute a habitat and be a true sphere, and concludes that energy harvesting alone would be a sufficient function to justify searches for such technology, a philosophy endorsed and amplified by \citet{GHAT1}.

\citet{GHAT1} and \citet{Teodorani14} also endorsed \citeauthor{dysonletters}'s perspective that searches for Dyson spheres would be unlikely by themselves to discover conclusive proof of extraterrestrial technology, but emphasized that the anomalies discovered would be good targets for more dispositive forms of SETI (such as searches for radio or laser communication), allowing those searches to be more focused. Indeed, \citeauthor{dyson60}'s approach to SETI has been generalized to include searches for most other kinds of extraterrestrial technology exclusive of deliberate communicative transmissions, in what has been called ``Dysonian SETI'' by \citet{Bradbury11} \citep[and ``artifact SETI'' by others, see also][]{WrightTaxonomy,WrightAdHoc}.

The first use of the term ``Dyson sphere'' in the literature is apparently due to \citet{kardashev64}, but since then the term has been occasionally reserved for the idea of a complete or nearly complete monolithic shell, with other configurations getting other names (such as a ``Dyson swarm''). Rather than fuss over correct and precise terms for what are, after all, entirely hypothetical objects, in this article we will use the well-known term ``Dyson sphere'' generically, to refer to any collection of artificial material around a star that produces significant amounts of waste heat, regardless of its specific geometry or purpose. This preserves the best known term for such material in a manner more or less consistent with \citeauthor{dyson60}'s original suggestion. That said, in what follows we will often implicitly invoke spherical symmetry, a common orbital radius, and common physical properties of the swarm components (at least roughly). Nonetheless, much of the analysis will extend to other configurations, as well, and so should still be useful as a way to guide observational searches for them. 

\subsection{Prior Literature: Theory}
\label{Sec:theoryoverview}

The concept of a Dyson sphere in popular culture actually predates \citet{dyson60}: \citeauthor{dyson60} credits the novel {\it Star Maker} \citep{stapledon} with the original idea. Indeed, in science fiction the idea usually manifests not as a swarm of small objects but as a single object of planet size or larger (a ``megastructure''). Perhaps the most famous appearance of a Dyson sphere in popular culture is in the {\it Star Trek:\ The Next Generation} television episode ``Relics,'' which depicts a  monolithic sphere with an entire biosphere on its inside surface (which makes sense only because artificial gravity is ubiquitous technology in the {\it Star Trek} universe). Megastructures of other configurations are also common in popular culture, for instance in {\it House of Suns} \citep{reynolds2008house} and perhaps most famously in the {\it Ringworld} series of novels \citep{Ringworld}, where the megastructure is a single, gigantic, spinning ring of impossibly strong material centered on a star, with the biosphere held on the interior surface by centrifugal forces.

A general theory of categorizing spacefaring species by their energy use was presented by \citet{kardashev64}, who described\footnote{The scale has been extended beyond \citeauthor{kardashev64}'s three types by many authors (e.g.\ Gray (2020)), with extensions to noninteger types usually referencing \citet{sagan73b} who suggested that the types be separated by factors of $10^{10}$ and normalized such that a Type 1 civilization consumes $10^{16}$ W. Such an extension necessitates the use of Arabic numerals instead of Kardashev's Roman numerals, which is appropriate. See \citet{GHAT2} and references therein, but note two inconsistencies in that work: in their Equation~(1) the power is normalized by 10 MW, but most or all other authors (including \citeauthor{sagan73b}) normalize to 1 MW, and the proper citation is \citet{sagan73b}, not \citet{sagan73a}.} ``Type {\sc ii}'' ``civilizations'' as those capable of commanding the entire energy output of their star. His approach was general, but many other analyses of the gravitational, radiative, and thermodynamic properties of Dyson spheres invoke specific geometries, purposes, energy generation schemes, or other activities for the Dyson spheres.  Some examples are studies of the gravitational dynamics of monolithic rings around stars \citep[][and references therein]{McInnes03,Rippert14}, the Harrop-Dyson satellite that exploits solar wind particles instead of photons \citep{Harrop10}, spheres with an inside surface temperature near 300K \citep{badescu95}, much hotter and smaller Dyson spheres that radiate in the optical \citep{Osmanov18}, analyses of very cold Dyson spheres \citep{Lacki16}, and partial shells used for stellar propulsion or energy extraction \citep{badescu2000,Badescu06}. Studies have also examined Dyson spheres around white dwarfs \citep{Semiz15}, neutron stars \citep{Osmanov16}, black holes \citep{Inoue11,Opatrny17}, and X-ray binaries \citep{Imara18}. 

There are many extensions of Dyson's idea that are difficult to cite, either because they appear as variations on a theme in science fiction or because their most formal description is in the gray literature. Two particularly notable examples of the latter are the discussions in Anders Sandberg's ``Dyson Sphere FAQ"\footnote{\url{https://web.archive.org/web/20190616230802/https://www.aleph.se/Nada/dysonFAQ.html}} and Robert J.\ Bradbury's description of ``Matrioshka Brains."\footnote{\url{https://web.archive.org/web/20090223093348/http://aeiveos.com:8080/~bradbury/MatrioshkaBrains/index.html} and\\ \url{https://web.archive.org/web/20080820083427/http://www.aeiveos.com:8080/~bradbury/JupiterBrains/index.html}} The former is a nice overview of the idea of Dyson spheres, and includes many of the topics discussed in this paper. It also includes, for instance, a discussion of how the outside of a monolithic Dyson sphere could be constructed to have Earth-like temperature and surface gravity, so provide a literal living surface \citep[a possibility also discussed in][footnote 7]{GHAT1}.   ``Matrioshka brains'' are a specific suggestion of how a set of nested Dyson spheres would be used to maximize the use of a star's luminosity to perform calculations.

A technological species that could build a Dyson sphere could also presumably spread to nearby star systems. If Dyson spheres are a generic phenomenon of such spacefaring life, then one might expect a galaxy with one Dyson sphere to have many more. Such a species that enshrouded all of its galaxy's stars would be ``Type {\sc iii}'' on \citeauthor{kardashev64}'s scale, although that term today is often used to refer to galaxy-wide species that fall somewhat short of that limit.

\citet{GHAT1} and \citet{GHAT2} provide a discussion of the possibility of galaxies filled with Dyson spheres and their observable consequences in broad terms. \citet{Lacki19} explored these consequences under a variety of more specific scenarios, and described the optical properties of galaxies in which specific masses of stars are preferentially ``cloaked.''

Interestingly, there is little or no discussion in the literature of the possibility of directly {\it imaging} a Dyson sphere. If such a sphere were sufficiently close to earth, it might be easily resolved at microwave, infrared, or optical wavelengths: a 1 au sphere around the closest stars to the Sun would subtend nearly an arcsecond on the sky. Indeed, we now have many images of protoplanetary and debris disks in reflected and emitted light, and such methods should be sensitive to Dyson spheres, as well.

\subsection{Prior Literature: Observations}
\label{Sec:obsoverview}

Though the technical feasibility of searches for Dyson spheres was noted early on \citep{dyson60,sagan66},
thorough surveys could not begin in earnest until the launch of {\it IRAS}, which provided the first mid-infrared survey of the sky with the aereal coverage and sensitivity necessary for such work. \citet{Slysh85} describes the first interpretations of the dataset, which confirmed \citeauthor{dyson60}'s prediction: there would be many bright infrared Galactic sources, and that the primary difficulty would not be detecting such sources but distinguishing them from natural sources.

Analysis of this data set proceeded slowly over the next 25 years. \citet{timofeev00} conducted a cursory search identifying a few objects with blackbody-like SEDs among the {\it IRAS} point sources. \citet{jugaku04} used the results of {\it IRAS} and near-infrared measurements to search for Dyson spheres, finding no good candidates among 384 nearby solar-type stars. \citet{carrigan09a} used the spectroscopic information from {\it IRAS} to examine sources across the sky for potential Dyson spheres, concluding that all four good candidates were likely distant red giants.

\citet{Teodorani14} outlined a more modern approach to the problem, using {\it Spitzer} as a more sensitive probe for infrared excesses and, following the suggestion of \citet{Arnold05}, {\it Kepler} to look for transiting megastructures. \citet{GHAT2} describe a program using the {\it WISE} all sky survey to search for Dyson spheres (and also other galaxies endemic with them), including a parameterization of their properties in terms of observables (the ``AGENT'' formalism after its five parameters, see Section~\ref{sec:observables}). \citet{GHAT4} outlined a general set of photometric anomalies in the light curves of stars that would be indicative of transiting megastructures, and identified one particular case study illustrative of the concept \citep[``Boyajian's Star'',][and references therein]{WTF,Wright16}. 

\citet{GHAT2} pointed out that unless virtually no starlight escapes a Dyson sphere, the flux decrement from blocked starlight is many times smaller than the infrared excess from the Dyson sphere's waste heat.  Nonetheless, \citet{Zackrisson15} recommended searching first not for infrared excesses, but for optically underluminous stars (i.e.\ stars with disparate trigonometric and spectroscopic parallaxes), and then following up with sensitive infrared measurements to search for the ``missing'' luminosity. Both \citeauthor{GHAT2} and \citeauthor{Zackrisson2018} emphasized the value of Gaia as a way to greatly improve search capabilities for Dyson spheres, since they would help distinguish Galactic sources (i.e.\ stars) from the far more ubiquitous extragalactic infrared sources (such as AGN); and since they would allow for underluminous stars to be more precisely identified.

The first observational search for galaxy-wide populations of Dyson spheres was that of \citet{annis99b}, who examined a sample of over 100 galaxies of similar distance to search for any that were optically underluminous, consistent with a significant amount of starlight blocked by Dyson spheres. \cite{GHAT3} used the {\it WISE} all-sky survey to search for resolved sources of extended infrared emission to search for populations of Dyson spheres by their waste heat, and \citet{Lacki16} used {\it Planck} data to put upper limits on galaxies with all of their starlight being reprocessed to very low temperatures. \citet{Zackrisson15} combined these strategies, searching for optically underluminous galaxies and following up with infrared measurements to identify the whether ``missing'' luminosity was being emitted there.  None of these searches resulted in particularly good candidates for Type {\sc iii} species.

\subsection{Purpose and Plan}

This article serves many functions. Much of the literature on Dyson spheres is scattered across journals spanning many disciplines. Sections~\ref{Sec:theoryoverview} and \ref{Sec:obsoverview} thus serve as a brief overview of some of that literature, including most of the observational work done to date.

Section~\ref{sec:stabilty} examines the (non-)stability of a monolithic sphere around a star.  Section~\ref{sec:Radiative_feedback} develops a general framework for analyzing the radiative interactions between the star and Dyson sphere, and  Section~\ref{shell} derives expressions for the special but illustrative case of a spherical shell of material.  Section~\ref{sec:observables} connects these parameters to the observable properties of the star plus Dyson sphere, including a particular example that illustrates the effects of stellar feedback on the star.  Section~\ref{sec:structure_feedback} discusses the still unaddressed problem of the effects of a Dyson sphere on a star's internal structure, and Section~\ref{sec:optima} discusses the maximum efficiency of a Dyson sphere and its optimum size and exterior temperature.

\section{Stability of a monolithic Dyson sphere} \label{sec:stabilty}

It is often asserted, as by \citet{dysonletters}, for instance, that monolithic Dyson spheres are ``impossible'' or ``unstable,'' but rarely is this carefully explained. Here we illustrate three ways in which they would not be stable. 

\subsection{Gravitational non-stability}

The first illustration of the non-stability of a monolithic Dyson sphere can be deduced from Gauss's Law \citep[as described, for instance, by][]{Harrop10}. Consider first the potential inside of a uniform, empty sphere whose material exerts a force following a radial inverse square law:
\begin{equation}
    \vec{F} = \frac{k\hat{r}}{r^2}
    \label{inversesquare}
\end{equation}
for instance because it is a charged insulator (in which case the electric field generated by a bit of surface with charge $q$ is $\vec{E} = q\hat{r}/r^2$) or massive (in which case the gravitational field generated by a bit of surface with mass $m$ is $\vec{g} = -Gm\hat{r}/r^2$).

It is an elementary exercise to show that Gauss's Law implies that inside the sphere there is zero field, and outside the sphere can be treated as a point source. Since it does not exert any force on particles inside itself, by Newton's First Law particles inside the sphere exert no force on it. 

\begin{figure}
    \centering
    \includegraphics[scale=0.4]{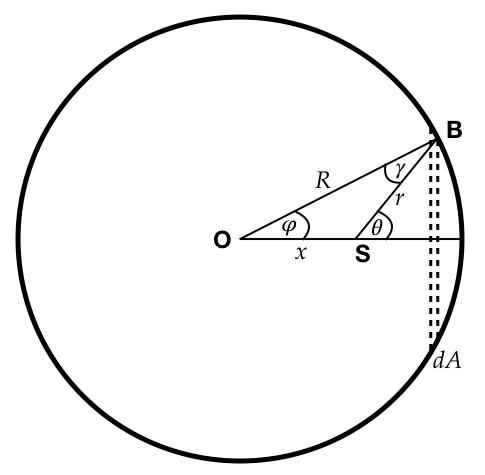}
    \caption{Figure after \citet{Osmanov18} illustrating the geometry of a particle within a spherical shell. The radius $R$ sphere is centered on $O$, with the particle at $S$ a distance $x$ from $O$ along the $x$ axis, and a distance $r$ from arbitrary point $B$ on the sphere. The circular strip having constant distance $r$ from the particle and  area $dA$ is indicated by dashed lines.}
    \label{fig:phi}
\end{figure}

This line of reasoning can appear somewhat abstract, and indeed it is not immediately obvious how well the result should generalize, for instance to a conducting sphere with induced charge generated by in interior point charge, to other closed shapes, or to surfaces intercepting interior sources of radiation.  Indeed, examining the geometry of \citet{Osmanov18} shown in Figure~\ref{fig:phi}, for $0>x>R$ it is not at all obvious that the combined force of the portion of the sphere to the right of point $S$ exactly cancels the force on that point from the other parts of the sphere.

And so as a didactic exercise, below I derive this result without invoking general theorems about vector fields.

The distance $r$ of arbitrary point on the sphere $B$ from the particle at point $S$ is 
\begin{equation}
\label{Eq:r}
    r = \sqrt{R^2 + x^2 - 2xR\cos{\varphi}}
\end{equation}
\noindent so, anticipating a change of variables we will perform later
\begin{equation}    
\label{Eq:dr}
    r dr = xR\sin{\varphi}d\varphi
\end{equation}

The area of an infintessimal circular strip of the surface containing B and centered on the $x$-axis can be written in terms of either $d\varphi$ or $dr$ as
\begin{equation}
    dA = 2\pi R^2 \sin{\varphi} d\varphi = \frac{2\pi R r dr}{x}
\end{equation}
and it feels a force from particle $S$ in the $x$ direction equal to
\begin{equation}
    dF_x = -\frac{CdA}{r^2} \cos{\theta} 
    \label{dFx}
\end{equation}
\noindent where $C$ depends on the nature of the force. By symmetry, the force in the $y$ and $z$ directions must be zero. Dropping a perpendicular onto the radius containing $\overline{OS}$ from $B$ and using Equation~(\ref{Eq:r}) gives
\begin{equation}
\label{Eq:thetaphi}
    r\cos{\theta} = R\cos{\varphi}-x = \frac{R^2-r^2-x^2}{2x}
\end{equation}
\noindent Then, we have for gravity or similar forces where $C$ is independent of the other variables, for all $|x|<R$ (and expressing things both in $d\varphi$ and $dr$ to illustrate the change of variable):
\begin{equation}
        F_x  = -2\pi C R^2 \int_0^\pi\frac{R\cos{\varphi}-x}{(R^2+x^2-2xR\cos{\varphi})^\frac{3}{2}} \sin{\varphi} d\varphi =-\frac{\pi C R}{x^2} \int_{R-x}^{R+x}\frac{R^2-x^2-r^2}{r^2}dr =0
\end{equation}

Thus, in the case of gravity a perfectly spherical and rigid sphere experiences no {\it net} force from a star interior to it, regardless of whether it is centered (but, as we will see, it will experience utterly destructive {\it differential} forces).  

This means that the centers of mass of the sphere and star are gravitationally uncoupled, and so there is no reason for the sphere to stay centered on the star---we can say it is {\it neutrally} stable in that it is in an equilibrium configuration that is neither stable nor unstable. 

Practically, the non-stability of a monolithic Dyson sphere means that it would require active station keeping to prevent any perturbative forces from causing it to drift into the star. The fact that it was only the symmetry of the sphere that led to this result also means that  deviations from sphericity would result in an unstable configuration.\footnote{For instance, if one corrugates part of the sphere, that patch will effectively have higher surface mass density than the other parts, and so feel a stronger force towards the star and thus break the symmetry that leads to zero force. Since the force will increase as the patch approaches the star, the configuration is locally unstable.}

\subsection{Radiation pressure non-stability}

If a sphere absorbs light from the central star, then radiation pressure acts as a force on any area element dA with a different force law, falling as $r^{-2}$ as with gravity or electromagnetism, but with an additional $\cos{\gamma}$ term to account for the projected area of the area element intercepting the photons.   We then have the force on the sphere in the x-axis from radiation pressure as:\footnote{\citet{Osmanov18} claim that this integral evaluates to a negative value, and that radiation pressure from the star will thus stabilize the sphere, but this is incorrect \citep[Z.\ Osmanov, private communication and][]{Osmanov19}.}
\begin{equation}
\label{Eq:radforce}
F_x = -2\pi C^\prime \int_0^\pi  \frac{R^2}{r^2}\cos{\gamma}\cos{\theta}\sin{\varphi}d\varphi
\end{equation}
\noindent where $C^\prime$ is independent of the other variables.

From Equations~(\ref{Eq:dr}) and (\ref{Eq:thetaphi}) we can change the variable of integration to $\theta$
\begin{equation}
    \frac{R}{r^2}\sin{\varphi}d\varphi = \frac{\sin{\theta}d\theta}{x\cos{\theta}+r}
\end{equation}
\noindent and dropping a perpendicular from $O$ onto the line containing $\overline{SB}$ gives
\begin{equation}
    x\cos{\theta}+r = R\cos{\gamma}
\end{equation}
\noindent so we have
\begin{equation}
 F_x= -2\pi C^\prime \int_0^\pi \cos{\theta}\sin{\theta}d\theta = 0
\end{equation}
\noindent and so even with the $\cos{\gamma}$ term, the sphere still feels no net force.

Remarkably, this result extends to {\it any} shape completely surrounding the star. To see this result another way, consider that by symmetry the ensemble of emitted stellar photons has zero momentum, and so any shape that absorbs all of them will not gain any momentum from them. Internal reflections or re-emission of these photons also will have no effect: because the photons are contained within the Dyson sphere they will inevitably strike another part of the Dyson sphere and deposit their momentum there. Indeed, the reflected photons as an ensemble can be thought of as purely internal thrust by the Dyson sphere against itself, and since such purely internal forces cannot change the center of mass of an object, the Dyson sphere cannot gain any net motion from this action and it remains neutrally stable to radiation pressure, no matter its shape.\footnote{Indeed, to the degree that this radiation is anisotropic (for instance from stellar flares or hot spots) the effect of this radiation is to destabilize the sphere.}

This exercise does reveal an interesting caveat, however: allowing stellar photons to escape from the Dyson sphere (or be preferentially re-emitted from the outside surface farther from the star) could provide thrust that would recenter the Dyson sphere on the star. Indeed, \citet{Shkadov} and \citet{Badescu06} describe ``stellar engines'' where exactly this action allows a partial sphere to not only keep itself centered, but to act as a gravitational tug and alter the trajectory of a star through the Galaxy.

\subsection{Lifetime of Thrust Stabilization}
\label{sec:thrust}

Active station keeping via thrust requires energy and propellant mass. \citet{Osmanov16} considered the energy requirements to stabilize a ring around a pulsar but here we consider the general case.

Because it is not dynamically stable, a monolithic Dyson sphere at equilibrium must be held there against the perturbative forces that would displace it. For scale, we will consider a monolithic Dyson sphere of arbitrary shape (so, perhaps a ring), mass $M$, of characteristic radius $R$ from the star. The Dyson sphere suffers perturbative forces due to other masses $m_p$ at closest approach distance $r$ from the Dyson sphere (for instance, a distant ice giant planet in the system). 

These accelerations will cause the Dyson sphere to distort and suffer differential acceleration with respect to the star in complex ways that depend on the geometry of the problem. To a rough order of magnitude, we can calculate the force to be balanced simply as 

\begin{equation}
    F_{\rm pert} = \frac{GMm_p}{r^2}
\end{equation}

\noindent with the caveat that the true force may be much smaller for many geometries, and $r$ may vary greatly across the different parts of the Dyson sphere. The Dyson sphere can cancel these accelerations by using some of its mass as thrust, expelled at velocity $v$, producing a force on the Dyson sphere
\begin{equation}
    F = \dot{M}v
\end{equation}
The power $P$ used to generate this thrust is
\begin{equation}
    P = \frac{1}{2}\dot{M}v^2
\end{equation}
\noindent and so the mass loss rate of the Dyson sphere is 
\begin{equation}
    \dot{M} = \frac{F^2}{2P}
\end{equation}
\noindent meaning that the Dyson sphere will have a lifetime of order
\begin{equation}
    \tau_{\rm life} = \frac{M}{\dot{M}} = \frac{2PM}{F^2}
\end{equation}
Equating this to the perturbative forces of other masses in the system via $F=F_{\rm pert}$ yields
\begin{equation}
    \tau_{\rm life} = \frac{2Pr^4}{MG^2m_p^2} =  \left(\frac{P}{L_\odot}\right)\left(\frac{r}{\rm au}\right)^4 \left(\frac{M}{M_{\jupiter}}\right)^{-1}\left(\frac{m_p}{M_{\jupiter}}\right)^{-2} \text{200 yr}
\end{equation}
\noindent where $M_{\jupiter}$ is the mass of Jupiter and where we expect that $P$ will be some (potentially very small) fraction of the total output of the star. Thus, for a monolithic sphere of mass $M_{\jupiter}$ to be stable for timescales long enough for it to be discovered (say, $\tau_{\rm life} > 10^7 \text{yr}$) there can be no perturbers of more than roughly $10^{-2} M_{\jupiter}$ within 1 au, and no gas giant planets within 10 au. It is worth reiterating that this estimate (and its scalings with $r$ and $R$) is extremely sensitive to the assumed geometry of the problem.

If there are no planets in the system because they were used to construct the Dyson sphere, then it seems stability via starlight-powered thrust could be maintained with sufficient active control for cosmic timescales. Alternatively, the Dyson sphere might use mass from the star as fuel to stabilize itself.

\subsection{Mechanical Instability}
\label{sec:mechanical}

As asserted by \citet{dysonletters} and shown quantitatively by \citet{Papagiannis85b}, the gravitational forces acting on a Dyson sphere are so extreme as to render the possibility of a real, rigid Dyson sphere supported by its own material strength dubious.

To see the difficulty, consider why an area element $dA$ does not accelerate towards the star under the force of gravity. In the case of a free-floating object, one can invoke centrifugal force from an orbit, but this is not possible for a monolithic sphere containing the star \citep[even for a rotating sphere, the poles will have zero centrifugal force, see][]{Covington91}.  

Instead, all of the area elements feel the same radial force, which would naturally cause the sphere to shrink and therefore generate lateral, compressive forces between the area elements.  Because of the curvature of the sphere, there is a small outward component of this compression force, which increases as the sphere shrinks until it is sufficiently large to counteract the star's gravity, and the sphere achieves equilibrium. In this formal sense, monolithic spheres thus appear mechanically stable.

Real structures are not infinitely rigid, however, and will deform if the forces on them exceed their elastic strength. For perfect spheres subject to uniform external pressure, collapse can occur for two reasons: when the material's yield strength $\sigma_y$ is exceeded because the external pressure exceeds
\begin{equation}
P_{\rm crit, y} = 2\sigma_y \frac{\Delta R}{R}
\end{equation}
\noindent where $\Delta R$ is the thickness of the sphere, or when buckling occurs above a critical pressure, given to within a cofactor of order unity by
\begin{equation}
    P_{\rm crit, b} \approx E\left(\frac{\Delta R}{R}\right)^2
    \label{buckling}
\end{equation}
\noindent where $E$ is the elastic (Young) modulus \citep{Papagiannis85b, buckle}. For monolithic Dyson spheres, we (presumably) are in the thin regime so Equation~(\ref{buckling}) (the elastic buckling condition) controls. The force on a Dyson sphere per unit area from the star's gravity is given by \begin{equation}
    P_{\rm grav} = \frac{GM_*\rho \Delta R}{R^2}
\end{equation} 
\noindent where $\rho$ is the density of the sphere material. This means that to avoid catastrophic buckling the sphere material must have an elastic modulus
\begin{equation}
    E \gtrsim \frac{GM_*\rho}{\Delta R}
\end{equation} independent of the radius of the sphere.  Plugging in some characteristic values yields
\begin{equation}
    E \gtrsim \frac{GM_\odot(1 {\rm g/cm}^3)}{10 {\rm m}} \approx 10^{13}\, {\rm GPa}
\end{equation}
\noindent which exceeds the elastic modulus of the strongest known material, carbyne, by nine orders of magnitude. Indeed, since $E$ is proportional to the strength of the atomic bonds of the material, and since the carbon-carbon bond is the strongest in nature\footnote{Invoking nuclear material as a building block only makes things worse: the density of such material is $10^{14}$ times higher, but its Young modulus is ``only'' of order $10^8$ GPa \citep{Scheuer81}, and so it is {\it nineteen} orders of magnitude too weak.} it appears impossible for a Dyson sphere of any size to support itself via elastic forces against gravity. 

One tempting solution that fails is to thicken the sphere and lower its density, perhaps to $100$ km and 0.1 g/cc, bringing the necessary strength of the material within a mere four orders of magnitude of the carbon-carbon bond. But for $R = 1$ au this would give the sphere a total mass of $\sim 1.4 {\rm M}_\odot$, and so the sphere's self-gravity would then become the dominant source of buckling and no further thickening would help. Adding struts or other internal bolsters is also no help, because the struts  will suffer the same difficulties as the sphere itself.

These problems are exacerbated by the requirement that the sphere be perfectly uniform and spherical---any defects or deviations from sphericity will create additional differential forces on the sphere. Such deviations could be large because there is a significant travel time for pressure waves around the sphere---even light takes almost an hour to travel around a circle with radius 1 au. The means that perturbations to the sphere will persist for times of order hours before they can be corrected via elastic forces.

One could imagine buttressing the sphere in some manner other than using high-strength material, perhaps with extremely strong magnetic fields that prevent the sphere from deforming, or by relieving the gravitational forces with electrostatic forces by charging the central star and sphere. While such a monolithic sphere might not be {\it literally} impossible to build, such a complex and fragile feat of engineering offers no obvious mechanical advantages over Dyson's original suggestion of a swarm of material.

A better solution is to make the sphere {\it thinner}, and so lessen the forces on the sphere via radiation pressure. For instance, the ratio of radiation forces to gravitational forces  $\beta\sim 1$ when the surface mass density of a patch of material normal to the Sun is
\begin{equation}
    \sigma \approx \frac{L_\odot}{4\pi GcM_\odot} \approx 0.8\, {\rm g/m}^2 
    \label{eq:leaf}
\end{equation}
\noindent which is comparable to the thinnest gold leaf. A monolithic Dyson sphere of this surface density is thus not mechanically unstable, although it is still dynamically neutral and it is not obvious what the utility of such a thin sphere would be (but see Sections~\ref{sec:cold} \& \ref{sec:radiation}). This surface density also serves as a lower limit on material bound by gravity to a star. 

Alternatively, one might imagine concentric spinning rings which use centrifugal forces to balance the otherwise destructive mechanical forces, but these rings would then essentially be a connected swarm of material orbiting the star in common orbits, and so need not be a monolith at all. 

\section{Radiative Feedback and Observable Consequences of a Dyson Sphere}

\label{sec:Radiative_feedback}
Having dispensed with the canard of a monolithic sphere, we now turn to the radiative interactions between a Dyson sphere and the central star, with a rough mental model along the lines of \citeauthor{dyson60}'s original conception (i.e.\ a swarm of orbiting material).

\citet{badescu95} presents the thermodynamics of Dyson spheres, including the feedback on the star, under certain restrictions on the useful work performed by the Dyson sphere, the amount of starlight allowed to escape, and the temperature of the Dyson sphere.  Here, we perform a more general analysis for more arbitrary kinds of Dyson spheres, although we will not consider Dyson spheres with strong angular asymmetries \citep[e.g.\ Shkadov thrusters,][]{Shkadov}.

We are interested in the ultimate fate of stellar photons emitted from the star ($*$) and thermal photons emitted by the Dyson sphere ($s$). We will denote the fraction of stellar photons that are ultimately absorbed by the star $f_{*,*}$ and those absorbed by the Dyson sphere $f_{*,s}$.  Similarly, we will denote the fraction of thermal photons emitted by the Dyson sphere and absorbed by the star $f_{s,*}$ and those absorbed by the Dyson sphere $f_{s,s}$.  Those photons that ultimately escape the system we denote $f_{*,e}$ and $f_{s,e}$.

These and the quantities below represent appropriate averages over wavelength. We will distinguish between ``thermal'' photons emitted by the Dyson sphere and ``stellar'' photons or ``starlight'' by using a subscript $t$ for the former. We will refer to the ``luminosity'' of an object in the usual astrophysical sense of total power, where the luminosity of the Dyson sphere includes emission from both the interior and exterior surfaces.

Table~\ref{tab:defs} lists the symbols we will use for this section.

\begin{table}
\begin{centering}
\begin{tabular}{c|p{0.85\textwidth}}
\hline
     Symbol & Definition (quantity is dimensionless unless otherwise specified in italics)   \\
     \hline
     $A$ & {\it Area} of the outer surface of the Dyson sphere\\
     $a$ & Bond Albedo of inner surface of the Dyson sphere to starlight \\
     $a_t$& Bond Albedo of inner surface of the Dyson sphere to thermal emission from the Dyson sphere \\
     $\alpha$ & Fraction of $\tilde{L}$ that does not escape the system as starlight, due to the Dyson sphere \\
     $B_\lambda$ & The Planck function, expressed as a {\it specific intensity} or {\it spectral radiance} \\
     $\gamma$ & Fraction of $\tilde{L}$ ultimately emitted and lost to space as thermal luminosity by the Dyson sphere \\
     $d$ & {\it Distance} to the Dyson sphere from Earth \\
     $e$ & Absorptivity of inner surface of the Dyson sphere to starlight\\
     $e_{\rm ext}$& Emissivity of the outer surface to thermal emission\\
     $e_t$& Absorptivity of inner surface of the Dyson sphere to thermal emission from the Dyson sphere\\
     $\epsilon$ & Energy generated by the Dyson sphere by means other than starlight collection, expressed as a fraction of $\tilde{L}$\\
     $\zeta$ & Fraction of thermal luminosity of the Dyson sphere emitted by the interior surface \\
     $\eta$ & Thermodynamic efficiency \\
     $f_{s,*}, f_{s,s}, f_{s,e}$ & Fraction of thermal emission from the Dyson sphere absorbed by the star, absorbed by the Dyson sphere, and lost to space\\
     $f_{*,*}, f_{*,s}, f_{*,e}$ & Fraction of starlight absorbed by the star, absorbed by the Dyson sphere, and lost to space\\
     $f_{{\rm int},e}$ & Fraction of thermal emission from the interior surface of the Dyson sphere lost to space \\
     $L$ & Total emergent {\it power} from the stellar surface \\
     $L_{\rm int}$ & Total emergent {\it power} from the interior (star-facing) surface of the Dyson sphere. \\
     $L_s$ & Total luminosity of the Dyson sphere. The sum of the thermal emission from the interior and exterior surfaces. {\it Power} \\
     $\tilde{L}$ & Luminosity of the star due to power generated in the stellar core. {\it Power} \\
     $n$ & Mean number of times a photon emitted from the interior of the Dyson sphere travels a chord across the Dyson sphere \\
     $\nu$ & Fraction of $\tilde{L}$ used by the Dyson sphere to do useful, non-dissipative work, such as emission of low entropy emission\\
     $r$ & Rate of computation of the Dyson sphere {\it Calculations per second}. \\
     $R$ & Radius of the Dyson sphere. {\it Distance}\\
     $R_*$ & Radius of the star.  {\it Distance}\\
     $S$ & {\it Entropy} (has units of Boltzmann's constant) \\
     $s$ & Probability a photon emitted by inner surface of the Dyson sphere does not immediately strike the star \\
     $\sigma$ & Stefan-Boltzmann constant\\
     $t$ & Transmittance of the Dyson sphere to starlight \\
     $t_t$ & Transmittance of the Dyson sphere to thermal emission from the Dyson sphere \\
     $T_{\rm eff}$ & Effective {\it temperature}, defined as the temperature of the blackbody generating a given luminosity with a given radiating area\\
     $T_*$ & Effective {\it temperature} of the star (used for brevity in Section~\ref{sec:optima}) \\
     $T_e$ & Effective {\it temperature} of the exterior of the shell (used for brevity in Section~\ref{sec:optima}) \\
     $T_{\rm min}$ & Minimum {\it temperature} of the Dyson sphere, set by the interstellar radiation field\\
     $\Phi_\lambda$ & {\it Specific flux} or {\it spectral irradiance} of light received at Earth
\end{tabular}
     \caption{Definitions of some symbols used in this document in sections \ref{sec:Radiative_feedback} and \ref{sec:optima}. Symbols in other sections may have different definitions.\label{tab:defs}}
     \end{centering}
\end{table}

\subsection{Special case of a spherical shell}
\label{shell}

\subsubsection{The fate of a photon leaving the interior of the Dyson sphere}

Here, we consider the special case of the fate of photons emerging from a star of radius $R_*$ centered in a sphere with radius $R$ and Bond albedo $a$. The sphere transmits a fraction $t$ of the starlight that reaches it, and has absorptivity $e$ such that 

\begin{equation}
    a+t+ e = 1
\end{equation}
\noindent (where we have chosen the symbol $e$ in anticipation of an application of Kirchhoff's law later).

From the perspective of the interior surface of the sphere, the star subtends a solid angle $2\pi(1-s)$, where
\begin{equation}
    s = \sqrt{1-(R_*/R)^2}
\end{equation}
\noindent represents the probability that a photon emitted from or reflected by the interior of the sphere
in a random direction will not strike the star before it strikes the sphere again.\footnote{In the formalism of \citet{badescu2000}, our $s$ is equivalent to their $\cos{\delta}$ or $\sqrt{1-x^2}$.}

This sphere is an approximate stand-in for any Dyson sphere, where $t$ may represent directions with no orbiting material. While our parameter $s$ invokes a single orbital radius for the material around the star, the formalism can likely be generalized for a swarm at a variety of distances by using some effective value for $s$. 

We consider an ensemble of photons emitted by the interior of the sphere, a fraction $s$ of which will immediately strike the star, and the rest of which will reach the interior of the sphere again. 

In the limit of purely specular reflection, a photon that misses the star and is reflected will necessarily miss the star on the second trip across the sphere, and so all of the remaining $1-s$ photons will continue to reflect until absorbed or transmitted, with probability $e$ and $t$, respectively each time they encounter the sphere.  The expectation value $n$ for the number of times a photon in this ensemble makes a chord across the sphere is thus 
\begin{flalign}
   && n = \sum_{i=0}^{\infty} a^i = \frac{1}{1-a} && \text{(specular case)}
   \label{specular}
\end{flalign}

In the limit of purely Lambertian scattering, a photon has a probability of striking the star of $(1-s)$ per trip, so by a similar calculation to above the expectation value $n$ is
\begin{flalign}
   && n = \sum_{i=0}^{\infty} a^is^i = \frac{1}{1-as} && \text{(diffuse case)}
   \label{diffuse}
\end{flalign}

One might approximate more general reflectors as a compromise between these two expressions for $n$.

\subsubsection{Fate of Stellar Photons}

The fate of stellar photons depends on the nature of the reflection from the inside of the sphere, given by the Bond albedo $a$.

In the case of specular reflection, by symmetry any stellar photon reflected by the sphere will return to the star, and so the fraction of starlight absorbed by the sphere is simply  $f_{*,s,{\rm spec}} =  e$, and the fraction  returned to the star is $f_{*,*,{\rm spec}} = a$. A fraction $f_{*,e,{\rm spec}}=t$ therefore escapes.

For diffuse reflection from a Lambertian inner surface of the sphere, a photon has a probability $a(1-s)$ of reflecting back to the star and $as$ of reflecting back to the sphere. The star thus absorbs a fraction $f_{*,*,{\rm diff}}=a(1-s)n$ of its own photons, the sphere absorbs a fraction $f_{*,s,{\rm diff}} =  e n$, and a fraction $f_{*,e,{\rm diff}} = tn$ is lost to space.

\subsubsection{Thermal emission of the sphere}

The radiative properties of the interior of the sphere depend on how it processes the stellar energy it does not reflect or transmit, and so it might reradiate any fraction of the starlight it absorbs as thermal radiation. Here we will denote the thermal luminosity emitted by the entire sphere $L_s$, and fraction of this emitted from the interior surface of the sphere
\begin{equation}
    \zeta = \frac{L_{\rm int}}{L_s}
\end{equation}

There is no reason that the effective reflection, absorption, or even transmittance properties of the sphere should be the same for this thermal emission as it was for the starlight, so we use the subscript $t$ to indicate the thermal versions of these properties.

For diffuse reflection, by a similar calculation to the one for the starlight we have that a fraction $f_{s,*,{\rm diff}} = \zeta (1-s)n_t$ is returned to the star---which is generally a small correction--- a fraction $f_{s,s,{\rm diff}} = \zeta s e_t n_t$ is returned to the sphere, and a fraction $f_{s,e,{\rm diff}} = \zeta st_t n_t$ of the interior luminosity is transmitted (with another fraction $1-\zeta$ radiated directly to space). 

In the case of specular reflection, the fate of these thermal photons is simple: a fraction $f_{s,*,{\rm spec}} = \zeta (1-s)$ strike the star, and the rest strike the sphere until they are either absorbed or transmitted.  Of the total thermal emission then, a fraction $\zeta s e_t n_t$ is ultimately absorbed, and a fraction $1-\zeta+\zeta st_tn_t$ is  radiated away. 

We combine the results of the fate of starlight and thermal emission of the sphere in the specular and diffuse cases in Table~\ref{tab:photons}.
\begin{table}
\setlength{\extrarowheight}{10pt}
    \centering
    \begin{tabular}{ccccc}
    & \multicolumn{2}{c}{{Starlight (*)}} & \multicolumn{2}{c}{{Thermal Emission from Sphere (s)}}\\[-10pt]
    & {Diffuse} & {Specular} & {Diffuse} & {Specular}\\
\hline
Absorbed by Star (*)  & $\dfrac{a(1-s)}{1-as}$ & $a$ & $\dfrac{\zeta(1-s)}{1-a_ts}$ & $\zeta(1-s)$\\
Absorbed by Sphere (s) & $\dfrac{ e}{1-as}$ & $ e$ & $\dfrac{\zeta s e_t}{1-a_t s}$ & $\dfrac{\zeta s e_t}{1-a_t}$   \\
Escape (e) & $\dfrac{t}{1-as}$ & $t$ & $1-\zeta + \dfrac{\zeta st_t}{1-a_t s}$ & $1-\zeta + \dfrac{\zeta st_t}{1-a_t}$\\[10pt]

\hline
    \end{tabular}
    \caption{Expressions for the fraction $f$ of stellar and thermal sphere photons that ultimately are absorbed by the star, absorbed by the sphere, or escape, for a spherical shell, in the limits of purely diffuse and purely specular reflection from the shell.}
    \label{tab:photons}
\end{table}

\subsection{Relating the Dyson Sphere Properties to Observables}

\label{sec:observables}

Following \citet{badescu2000}\footnote{In what follows we roughly follow the notation of \citet{badescu2000}, however readers wishing to compare the two treatments should note that we use the subscript $s$ to denote the sphere, while they use the same symbol to denote the {\it star}. We also use $a$ to represent the Bond albedo, where they use $a$ to represent the {\it absorptance} (which we refer to as e).}, we refer to the intrinsic luminosity of the star from all interior processes (e.g.\ nuclear burning) as $ \widetilde{L}$, distinguished from $L$, the luminosity from its surface, which is somewhat higher than $ \widetilde{L}$ because some if its emission is being returned by the sphere, and the emergent flux on the surface must increase to maintain energy balance. 

Following the AGENT formalism of \citet{GHAT2}, we consider the energy budget of the star plus sphere, where the sphere absorbs a fraction $\alpha$ of power $ \widetilde{L}$ generated by the star and generates power by other means equal to $ \widetilde{L}\epsilon$. The total luminosity of the system is then $ \widetilde{L}(1+\epsilon)$. The sphere ultimately radiates away to the external universe $ \widetilde{L}\gamma$ as thermal emission (which is lower than $L_s$ because some emission is returned to the sphere or star), and has a nonthermal luminosity $ \widetilde{L}\nu$ (for instance powerful laser or radio transmissions). By energy balance,\footnote{The sphere cannot do significant amounts of work on long timescales without ultimately converting the energy to a low entropy form such as nonthermal emission or as mass \citep[captured by our parameter $\nu$ and subject to the usual Carnot efficiency limits, but see][]{Badescu2014} or dissipating it as heat (captured by our parameter $\gamma$.) While other sinks of energy exist (for instance it could be stored as chemical or gravitational potential energy), they are infinitessimal compared to the total power emitted by the star over timescales on which such spheres must exist in order for us to detect one, so we can ignore them. See \citet{GHAT2}.} we then have 
\begin{equation}
    \alpha + \epsilon = \gamma + \nu
\end{equation}

For simplicity we will assume that any nonthermal emission is deliberately directed away from the star. By energy balance on the surface of the star we have
\begin{equation}
   \widetilde{L} + L_sf_{s,*} = Lf_{*,s} + L f_{*,e} 
\end{equation}
\noindent and on the sphere we have
\begin{equation}
     \widetilde{L}\epsilon + L f_{*,s}  = L_s f_{s,*} + L_s f_{s,e} +  \widetilde{L}\nu
\end{equation}
Solving these above two equations for $L$ and $L_s$ we have
\begin{eqnarray}
    \frac{L}{ \widetilde{L}} = \frac{(1-f_{s,s}) + f_{s,*}(\epsilon-\nu)}{f_{s,*}f_{*,e} + f_{s,e}(1-f_{*,*})}\\
    \frac{L_s}{ \widetilde{L}} = \frac{f_{*,s}+(1-f_{*,*})(\epsilon-\nu)}{f_{s,*}f_{*,e}+f_{s,e}(1-f_{*,*})}
\end{eqnarray}

Then we can express the two AGENT parameters that describe the observability of the system, $\alpha$ and $\gamma$, as
\begin{equation}
    \alpha = 1 -\frac{L}{ \widetilde{L}}f_{*,e}
    \end{equation}
    \begin{equation}
    \gamma = \frac{L_s}{ \widetilde{L}}f_{s,e}
\end{equation}

The effective temperature of the star is then
\begin{equation}
    T_{*,{\rm eff}} = \left(\frac{L}{4\pi\sigma R_*^2}\right)^\frac{1}{4} = \left(\frac{L}{ \widetilde{L}}\right)^\frac{1}{4} \widetilde{T}_{*,{\rm eff}} \\
\end{equation}
\noindent and, if we choose a characteristic distance for the components of the sphere $R$, we can also define an effective temperature for its interior and exterior surfaces:
\begin{flalign}
    T_{{\rm int, eff}} &= \left(\frac{\zeta L_s}{4\pi\sigma R^2 e} \right)^\frac{1}{4} = \left(\frac{\zeta L_s}{ e  \widetilde{L}}\right)^\frac{1}{4} \widetilde{T}_{s,{\rm eff}}\\
    T_{{\rm ext, eff}} &= \left(\frac{(1-\zeta) L_s}{4\pi\sigma R^2 e_{\rm ext}}\right)^\frac{1}{4} = \left(\frac{(1-\zeta) L_s}{ e_{\rm ext}  \widetilde{L}}\right)^\frac{1}{4} \widetilde{T}_{s,{\rm eff}} 
\end{flalign}
\noindent where we have applied Kirchhoff's law to approximate the emissivity of the inner shell surface by its absorptivity $e$, and we have introduced $ e_{\rm ext}$, the emissivity of the external surface of the sphere, which up to now has not been relevant to the problem, and is only constrained to be $ e_{\rm ext}<1-t_t$. We have also introduced nominal values of the effective temperature of the star (the one it would have without the sphere around it) and the sphere (the value it would have if it had no feedback on the star and all of the flux was radiated on the outside surface):
\begin{align}
     \widetilde{T}_{*,{\rm eff}} & \equiv \left(\frac{ \widetilde{L}}{4\pi \sigma R_*^2}\right)^\frac{1}{4}\\
     \widetilde{T}_{s,{\rm eff}} &\equiv \left(\frac{ \widetilde{L}}{4\pi \sigma R^2}\right)^\frac{1}{4} \label{Tseff}
\end{align}

We can then express the total spectrum of the system as three components: the stellar spetrum, the interior of the shell, and the exterior of the shell.  The specific flux received at Earth $\Phi_\lambda$ depends on the distance to the star, and for spherical blackbody radius $R$ at temperature $T$ is given by
\begin{equation}
    \Phi_{\lambda,{\rm BB}} = \pi B_\lambda(T) \frac{R^2}{d^2}
\end{equation}
where $B_\lambda$ is the Planck function and $d$ is the distance to the object. 

For the star-Dyson sphere system, a fraction $f_{*,e}$ of the star's emission escapes. The total emission from the sphere lost to space is $L_sf_{s,e}$, divided between the exterior surface, emitting a fraction $(1-\zeta)$ of the {\it total} sphere luminosity $L_s$, and the interior surface, emitting the balance, a fraction $f_{{\rm int},e}$ of which escapes. Solving for $f_{{\rm int},e}$ we have
\begin{eqnarray}
    L_s f_{s,e} &=& \zeta L_s f_{{\rm int},e} + (1-\zeta) L_s \\
    f_{{\rm int},e} &=& (f_{s,e} - (1-\zeta)) / \zeta
\end{eqnarray}

The spectrum of the specific flux received at Earth from the star-Dyson-sphere system is then
\begin{equation}
    \Phi_\lambda = \frac{\pi}{d^2} \left(
    k_1 B_\lambda(T_{*,{\rm eff}}) +
    k_2 B_\lambda(T_{{\rm int,eff}})  +
    k_3 B_\lambda(T_{{\rm ext,eff}})
    \right)
\end{equation}
\noindent where we have approximated the spectra of the three components by the Planck function for illustrative purposes and where
\begin{eqnarray}
    k_1 &=& R_*^2 f_{*,e}\\
    k_2 &=& R^2 e f_{{\rm int},e} \\
    k_3 &=&  R^2 e_{\rm ext} 
\end{eqnarray}

If a Dyson sphere were discovered, and if it conformed to the assumptions of this analysis, then the effective temperatures and relative strengths of the three components combined with a parallactic distance would thus directly reveal information about the properties of the Dyson sphere, parameterized here by $R, a, e, a_t, e_t, e_{\rm ext}, \epsilon, \nu$ and $\zeta$, and the star, parameterized by $\tilde{L}$ and $R_*$. This is eleven parameters to be deduced from, at most, 6 observables, leaving the system underconstrained. Further, it is unlikely that the interior and exterior components of the sphere's emission would be observationally distinct, since the thermal emission will likely be dominated by one of them. Analysis would need to proceed by making assumptions about these parameters such as $\epsilon \ll 1$ and $\nu \ll 1$, $\eta \sim 0.5$, $e_{\rm ext} = e$, and to constrain $\tilde{L}$ and $R_*$ using knowledge of stellar structure (but see Section \ref{sec:structure_feedback}).

In reality, a Dyson sphere may have materials at a range of distances with a variety of properties, and a more informative spectrum that can be better modeled. But this parameterization allows for connecting upper limits and Dyson sphere candidates to the regions of parameter space excluded or allowed by observations.

\subsection{Example of Feedback from a Dyson Sphere}
As an illustration of these results, we make use of the results of Section~\ref{shell} to consider the simplified case of a passive spherical shell ($\nu = \epsilon = 0$) where the properties of the sphere are constant on the interior and exterior ($\zeta = 0.5$) and between starlight and thermal emission (i.e.\ dropping the $t$ subscript). In the case of an even mixture of specular reflection, absorption, and transmittance ($a=t= e= e_{\rm ext} = \frac{1}{3}$), we have 
\begin{eqnarray}
    \frac{L}{ \widetilde{L}} &=& \frac{4-s}{2}\\
    \frac{L_s}{ \widetilde{L}} &=& \frac{2}{3}\\
    \alpha &=& \frac{s+2}{6}\\
    \gamma &=& \frac{s+2}{6}\\
    T_{*,{\rm  eff}} &=&  \widetilde{T}_{*,{\rm  eff}} \left(\frac{4-s}{2}\right)^\frac{1}{4}\\
    T_{{\rm int, eff}} &=&  \widetilde{T}_{s,{\rm eff}}\\
    T_{{\rm ext, eff}} &=&  \widetilde{T}_{s,{\rm eff}}
\end{eqnarray}

\begin{figure}
    \centering
    \includegraphics{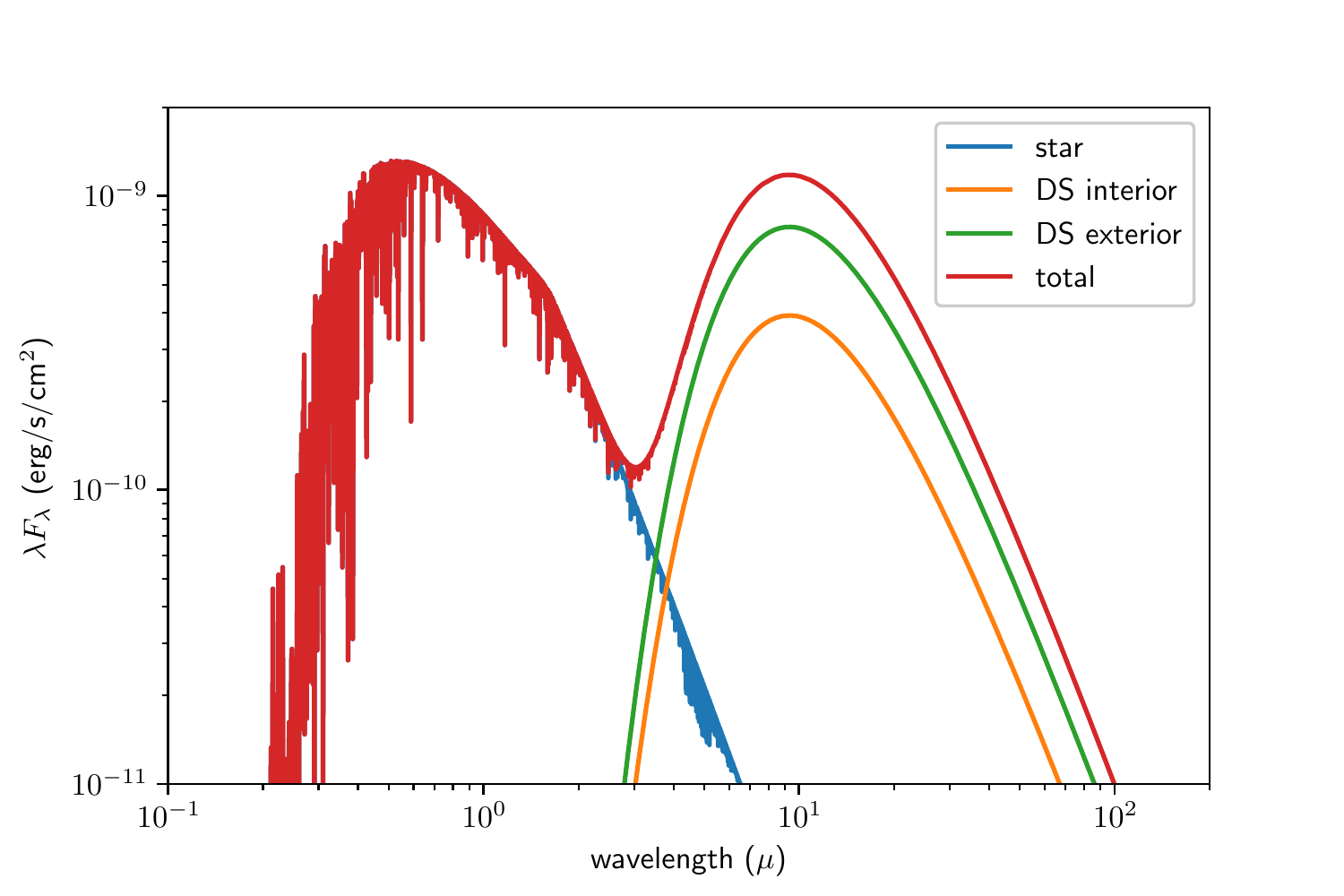}
    \caption{Spectrum of a Sun-like star surrounded by a 1 au Dyson sphere, as observed from a distance of 100 pc.  The sphere transmits 1/3, absorbs 1/3, and reflects 1/3 of the photons that reach it, and radiates waste heat evenly between its interior and exterior surfaces, which have the same effective temperatures. The interior surface is partially obscured by the sphere itself and so appears dimmer. The stellar spectrum blueward of 24$\upmu$ is a $T_{\rm eff}=5,800$ K dwarf star spectrum from the NextGen stellar atmosphere grid \citep{Allard97,NEXTGEN}.}
    \label{fig:spectrum}
\end{figure}

Figure~\ref{fig:spectrum} shows the spectrum of such a system where $R$ = 1 au, and the central star is Sun-like, as observed at a distance of 100 pc. 

More generally, in these equations we can see the effects of feedback of the sphere on the star. In the limit of a very large sphere ($s\rightarrow1$) $1/3$ of the starlight is reflected back onto the star, where it must ultimately reemerge as starlight, again with $1/3$ reflecting back on the star. This series converges on a value of $1/2$, meaning the stellar surface has a luminosity 50\% higher than the core ($1/3$ of which never leaves the sphere). 

Meanwhile, $f_{s,s}=1/4$ of the emission from the sphere is trapped within the sphere, and the remaining $f_{s,e}=3/4$ that escapes is powered by the $L/3$ of starlight it absorbs. Its total luminosity is thus $2 \widetilde{L}/3$, but only $ \widetilde{L}/2$ ultimately emerges as waste heat.

The effective temperature of the star is somewhat higher to accommodate the extra surface luminosity. The effective temperatures of the surfaces of the sphere in this case are just as we would expect from a naive calculation without stellar feedback, however.

The picture is altered for Dyson spheres smaller than a few stellar radii, because then the star captures a significant amount of the sphere's waste heat. This reprocesses the waste heat back into starlight, lowering $\alpha$ and $\gamma$. In the limit that the sphere sits on the surface of the star ($s\rightarrow0$) it simply blocks its geometric area and we have $\alpha=  \gamma=1/3$. In this case the total surface luminosity of the star grows to $2 \widetilde{L}$, $1/3$ of which is reflected by the sphere, and $1/3$ of which is absorbed by the sphere.  Of the $1/3$ that is absorbed, half escapes to space and the other half is returned to the star. The star thus has $2/3  \widetilde{L}$ escape as starlight and $1/3  \widetilde{L}$ reprocessed as waste heat.

\subsection{Feedback on Stellar Structure}

\label{sec:structure_feedback}

Up to this point we have assumed that the luminosity of the stellar interior, $\widetilde{L}$, is not affected by the presence of the Dyson sphere, but this is not strictly true. To first order, small amounts of energy returned to the stellar surface will be re-emitted as starlight (the ``reflection effect'' of stellar binaries) but if the energy is carried into the stellar interior by convection or other means, then it should alter the structure of the star, and therefore its total luminosity (not just the luminosity of the surface) on a Kelvin-Helmholz timescale.

To see this, note that a key feature of the solutions to the equations of stellar structure is that stars have negative {\it gravitothermal heat capacity}. This is a common feature of gravitationally bound systems: adding energy to them causes them to expand and cool (i.e.\ their components slow down). This is a consequence of the Virial theorem, and is akin to adding energy to a satellite: the satellite will attain a higher, less bound orbit, and as a result move more slowly.

Similarly, stars are supported against gravity by their thermal energy, which is ultimately provided by nuclear burning. When a Dyson sphere returns some of the escaping energy to the star, this provides additional support against gravity, and in response the star should either expand, lower its luminosity, or both. In fact, we expect both since the pressure on the core is roughly (to an order of magnitude)

\begin{equation}
    P_c \approx \frac{GM^2}{R_*^4}
\end{equation}

\noindent and the luminosity of the star is extremely sensitive to central pressure. This is the thermostat that keeps the core temperature constant, and keeps stars stable. Without such a thermostat, any perturbative increase in core luminosity would lead to an increase in core temperature and pressure, which would lead to higher luminosity in a positive feedback loop, and every star would be destroyed in a runaway thermonuclear explosion, or cool and collapse until held up by electron degeneracy pressure.\footnote{If the star is shell-burning, the shell's pressure and therefore luminosity is set mostly by the surface gravity of the degenerate core, not the weight of the envelope above it. The envelope then expands in response to the growing core luminosity without negative feedback, and the star becomes a giant. In this case, the luminosity of the star is regulated by the lifting of the degeneracy of the core at high temperatures, which causes the core to expand, its surface gravity to drop, and the shell luminosity to decrease.}

In principle, therefore, a sufficiently insulating Dyson sphere could put enough energy back into its central star that the star might expand, cool, and dim. This might be done intentionally, for instance to prolong the life of a star and prevent it from entering the giant phase, or it might be an unwanted consequence of consuming most of the star's output.  Regardless of the motivations for initiating such feedback, however, the effect on the star will be a calculable consequence of the properties of the Dyson sphere, meaning that it is possible to connect observables like luminosity and effective temperature of the starlight and waste heat to the physical parameters of the Dyson sphere.

This is an area of study ripe for investment, since the overall effects of Dyson spheres on the total luminosity of stars have not previously been studied, and might lead to strong observable consequences that strengthen their detectability in current or near-future all-sky surveys. 

\section{Efficiency and Optimal Size of a Dyson Sphere}
\label{sec:optima}

\subsection{Assumptions}

Some authors \citep[e.g.][]{Suffern77,badescu95} have presumed that Dyson spheres would constitute living surfaces (either on the inside, outside, or in between) and/or that they would be engineered to maximize the work that could be done or the efficiency with which starlight could be processed. In this work, we have generally remained agnostic with respect to the purpose or design of a Dyson sphere. Here, we will follow the spirit of \citet{Dyson66} and explore the maximum efficiency of a Dyson sphere allowed by physics as a function of its size, with minimal appeal to engineering and function. 

For scale, we will refer to nominal numbers for a ``canonical'' Dyson sphere consisting of a complete shell of orbiting material at $R=1$ au from a Sun-like star with $T_*= 5772$ and $R_*= R_\odot$.  To be useful, the elements of the shell must have some substance, which requires a large amount of mass.  If the density of the shell is of order 1 g/cc, then one Jupiter mass would provide for components a few meters thick. 

We will make the assumption that the star is its only important source of energy (i.e.\ $\epsilon = 0$). Maximum efficiency implies that all of the starlight is captured, and that the Dyson sphere does no work on the star, so we are in the limit where $e=e_{\rm ext}=1$, $f_{*,s}=1$, $f_{s,*}=0$, $L=\tilde{L}$, and the total waste heat of the Dyson sphere $L_s f_{s,e} = L (1-\nu)$.

If Dyson spheres are common enough to be observed by us, then they must be long-lived, so the nature of their work must be sustainable on cosmic timescales.\footnote{This assumes that we do not live in a special cosmic time in which the rate of Dyson sphere creation is suddenly sharply rising above zero. Some have questioned this assumption, most notably \citet{annis99a} and \citet{cirkovic08} who have suggested that a recent ``astrobiological phase transition'' in the Galaxy can help explain the Fermi Paradox \citep[but see][]{Carroll19}.} \citet{GHAT2} argued that the long-term nature of the work done by a Dyson sphere will likely be dissipative or emissive: the star's energy cannot be stored on cosmic timescales without heating or unbinding the Dyson sphere (except, perhaps, via energy-to-mass reaction). This is illustrated by the Kelvin-Helmholtz timescale for the Sun: the Sun generates enough energy to completely unbind itself in $\sim 30$ Myr. The gravitational binding energy of the planets is significantly smaller than this: the Sun generates enough energy to unbind every atom in Jupiter and eject it from the Solar System in less than 1,000 years. Only if the work of the Dyson sphere is to deconstruct the star should we expect to find it doing lasting work on its system.

Whatever work a Dyson sphere does must therefore result in the energy leaving the system or being stored in some manner with significantly more capacity than heat or gravitational potential energy. The only obvious such mechanism would be to store the energy as mass. While converting starlight into, say, protons would seem inefficient (collecting mass from the stellar wind would be at least as effective), the production of antimatter is plausible activity that would result in little emission. 

We then have two limits to consider: that all of the work done by the Dyson sphere goes into the creation of mass or the emission of low-entropy radiation (either way parameterized by $\nu$ in the AGENT formalism above) or else is used in a dissipative way and emitted as waste heat (as in, for instance, computation).

\subsection{Maximum Thermodynamic Efficiency of Dyson Spheres}

If the purpose of the Dyson sphere is to do work, the thermodynamic efficiency $\eta$ with which it does this work can be computed as the fraction of starlight that goes into this work, equivalent to the starlight it receives minus the waste heat the Dyson sphere disposes of:

\begin{eqnarray}
    \eta &=&  \frac{L-4\pi R^2 \sigma (T_{\rm{ext,eff}}^4-T_{\rm min}^4)}{L} \nonumber \\
    &=& 1 - \frac{R^2(T_e^4-T_{\rm min}^4)}{R_*^2T_*^4} 
\end{eqnarray}

\noindent where for brevity we use $T_e \equiv T_{\rm ext,eff}$ and drop the tilde and eff subscript on the stellar effective temperature and where we have, for the first time in this work, included the effects of outside radiation falling onto the Dyson sphere, which would approach $T_{\rm min}$ in the absence of heating from the star.

The value of $T_{\rm min}$ will depend on the Dyson sphere's environment, but typical values of the interstellar radiation field near the Sun are of order 1 eV/cm$^3$ \citep{Mathis83,Strong2000}, corresponding to $T_{\rm min} \sim 4$ K.

\citet{Badescu2014} explores the nuances of calculating thermodynamic efficiency when the source and sink of an engine is radiation instead of a thermal bath. Following the spirit of that work,\footnote{\citeauthor{Badescu2014}'s treatment is not directly applicable to our problem because all of the thermal photons radiated from the interior of our Dyson sphere strike the Dyson sphere again while \citeauthor{Badescu2014} implicitly assumes all of those photons return to the heat source.} we consider the entropy received and emitted by the Dyson sphere per unit time:

\begin{eqnarray}
    \dot{S}_{\rm in} &=& \frac{4}{3}4\pi\sigma\left(R_*^2T_*^3 + R^2T_{\rm min}^3\right) \label{eq:sin}\\
    \dot{S}_{\rm out} &=& \frac{4}{3}4\pi\sigma R^2T_e^3
    \label{eq:sout}
\end{eqnarray}

\noindent The Dyson sphere works at maximum efficiency when the work it does generates no entropy, meaning 

\begin{equation}
    \Delta \dot{S} \equiv \dot{S}_{\rm out}-\dot{S}_{\rm in} = 0
\end{equation}

\noindent yielding

\begin{equation}
    \left(\frac{R_*}{R}\right)^2 = \frac{T_e^3-T_{\rm min}^3}{T_*^3}
\end{equation}

\noindent Applying this constraint, we find that a conservative upper bound of thermodynamic efficiency for a Dyson sphere of temperature $T_e$ of:

\begin{equation}
    \eta < 1 - \frac{T_e}{T_*}\frac{1-(T_{\rm min}/T_e)^4}{1-(T_{\rm min}/T_e)^3}\approx 1-\frac{T_e}{T_*}
   \label{Eq:eta}
\end{equation}

\noindent where the latter approximation applies to waste heat temperatures significantly above $T_{\rm min}$, and is equivalent to the Carnot limit. More realistic Dyson sphere geometries and properties will yield lower efficiencies.  This expression is simply a measurement of the energy available for work (i.e.\ the {\it exergy}\footnote{{\it Exergy} is thermodynamics jargon for the theoretical limit of the amount of energy available in a system for it to do work, given its environment. It is equivalent to the Gibbs free energy for chemical systems at constant pressure or the Helmholz free energy for chemical systems at constant volume.}) that can be extracted by the Dyson sphere, given by the contrast between the (low) entropy of the energy it collects from the starlight (at $T_*$) and the (higher) entropy of its emitted radiation (at $T_e$).  

In the limit of infinitely large Dyson spheres, the temperature of the Dyson sphere $T_e \rightarrow T_{\rm min}$ and the two sides of Equation~\ref{Eq:eta} reach their maximum value: 

\begin{equation}
    \eta_{\rm max} = 1-\frac{4}{3}\frac{T_{\rm min}}{T_*}
\end{equation}

\noindent which is 99.9\% for our nominal Dyson sphere. Dyson spheres can, in principle then, convert most of the energy in starlight into work.

\subsection{Thermodynamic Efficiency of Low-Entropy Emission}

We now risk speculating on the {\it practical} limits of alien technology by considering the engineering optimum for such a Dyson sphere, as opposed to this theoretical limit. 

By energy balance we have that the relationship among the total radiating area $A$ of a Dyson sphere, its effective temperature, and its waste heat luminosity is simply

\begin{equation}
    L_s f_{s,e} = L (1-\nu) + A \sigma T_{\rm min}^4 = A \sigma T_e^4
    \label{eq:sigmaT4}
\end{equation}

\noindent If all of the work goes into low entropy emission or mass generation then we have $\nu = \eta$, and neglecting the background radiation for now, we find

\begin{equation}
    A = 4 \pi R_*^2 (1-\eta)^{-3}
    \label{eq:efficiency_nu}
\end{equation}

For reference, our nominal 1 au Dyson sphere would have $T_e=160$ K, and maximum efficiency $\eta = 97\%$. This is very close to the theoretical limit, though of course a species bent on capturing that last 3\% of exergy could do so with a larger Dyson sphere, at the expense of significantly higher engineering difficulty. 

Equation~(\ref{eq:efficiency_nu}) shows that for every additional ``nine" of efficiency, the Dyson sphere requires three orders of magnitude more radiating material. This implies that capturing another 2.5\% of the exergy of starlight by building a larger Dyson sphere of outer temperature 16K requires either a thousand of Jupiter masses (i.e.\ 1M$_\odot$) or making do with a Dyson sphere only millimeters thick. The latter option would seem to be at or beyond the limit of utility: even if the only function of the Dyson sphere is to serve as a radiator of waste heat, transporting that waste heat to the shell itself requires material, implying that the mass required for a Dyson sphere inevitably shares this steep scaling with efficiency (but see the caveats in Sections~\ref{sec:cold} \& \ref{sec:radiation}). But regardless of the details, unless a technological species somehow incurs negligible cost when adding material to its Dyson sphere, the optimum size of the Dyson sphere it builds will be significantly smaller than its theoretical limit.

\subsection{Thermodynamic Efficiency of Dissipative Work}

If the purpose of the Dyson sphere is not to emit large amounts of low entropy emission or create mass, then the work it does must ultimately be radiated away as waste heat, which increases the area of radiators needed (or the temperature of the waste heat). By the same calculation as above but with $\nu=0$ in Equation~(\ref{eq:sigmaT4}) to account for this extra energy to be radiated, we then have

\begin{equation}
    A = 4 \pi R_*^2 (1-\eta)^{-4}
\end{equation}

\noindent which is an even steeper function implying even smaller optimum Dyson spheres. In this case our nominal 1 au Dyson sphere has $T_e = 390$K and efficiency $\eta = 93\%$, and achieving 99\% efficiency with a larger Dyson sphere at the same surface mass density would require 2.4 M$_\odot$ of material.

\subsection{Efficiency of Computation}

In this latter case of purely dissipative work, one might follow \citet{Sandberg16} and consider computations the nominal kind of work performed since the amount of such work that can be done scales with exergy and does not result in any stored or emitted energy---an ideal function for a Dyson sphere. Indeed, \citet{Sandberg99} provides an extensive discussion of the limits of many aspects of Dyson sphere computation including memory storage density, computational speed, and energy use.

Briefly, Landauer's principle sets the minimum amount of heat generated by a single binary logical operation such as AND or NOR at $kT\ln{(2)}$. This has the nice consequence that colder computers can do more operations with a given amount of energy.  This means that large Dyson spheres win twice when their work is computational: once because they can extract more exergy from starlight, and again because they need less energy per computation.  

The rate of classical computation $r$ at a distance $R$ from a star at power level $L$ at $T_e$, and is

\begin{equation}
    r = \frac{\eta L}{kT_e\ln{(2)}}
\end{equation}

\noindent which for our nominal Dyson sphere is approximately $10^{47}$ logical operations per second.  This rate is maximized for very large Dyson spheres with $T_e \rightarrow T_{\rm min}$, which for the Sun would yield $10^{49}$ logical operations per second. 

We can then define a {\it computational} efficiency $\eta_{\rm comp}$ calculated as a Dyson sphere's computational rate compared to the maximum possible rate at $T_e\rightarrow T_{\rm min}$. 

\begin{equation*}
    \eta_{\rm comp} =\frac{r}{r_{{\rm max}}} = \frac{\eta}{\eta_{\rm max}}\frac{T_{\rm min}}{T_e}
\end{equation*}

\noindent which we have seen for our nominal Dyson sphere is around 1\%. It would seem, then, that when optimizing for computational rate instead of work, there might be great profit in investing in larger Dyson spheres.\footnote{Or in waiting long enough that $T_{\rm min}$ falls. \citet{Sandberg16} suggest that the reason we do not see Dyson spheres is that the would-be builders are ``aestivating,'' i.e.\ in a state of suspended animation until the Universe cools enough for their computers to run more efficiently. See, however, \citet{Bennett19} for a critique of this idea.} 

For Dyson spheres far from their limits (i.e.\ neglecting outside radiation and with $\eta\sim 1$), we can write $L\approx A\sigma T_e^4$, and so the scaling of computational rate with Dyson sphere area is 

\begin{equation}
    A = \frac{(k\ln{(2)})^4}{L^3 \sigma}r^4
\end{equation}

The cost function here is not as steep as with thermodynamic efficiency until $T_e\rightarrow T_{\rm min}$, but it still requires a factor of 16 in material to double computing power. Dyson spheres used for computing might thus have larger optimal sizes and lower optimal temperatures than other Dyson spheres, but there are still strongly diminishing returns. 

We use the qualifier ``classical'' to describe this computing because it remains unclear if the entropy produced by logic gates represents a fundamental limit of computing, or whether reversible computing techniques can surpass those limits \citep[see, e.g.][]{Frank02}. If so, then it is not immediately clear which aspects of Dyson spheres would optimize such computation, although \citet{Sandberg99} provides some possibilities.

For instance, all physical systems are subject to noise and error, and so the entropy generated (and energy disposed of) during error correction is likely a universal feature of all computers. A fully reversible computer might do many more logical operations with a given amount of energy than its classical counterpart, but it still needs a source of low entropy power to clear its computer memory for error correction. In this case, the rate of memory clearings a Dyson sphere can perform for a given exergy is equal to its classical computation rate $r$, and this might drive a similar design optimization.

\subsection{Very Cold Dyson Spheres}
\label{sec:cold}

While the cost function for large Dyson spheres is very steep, it is possible that a species might find ways to make the coefficient so small and the benefit function so large that very cold Dyson spheres would be common. For instance, if the purpose of the Dyson sphere was only to shift starlight to the lowest possible frequency, and not to do any useful work, then many of the above considerations are not relevant. In this case the minimum surface density allowed by physics is quite low, because its only function is to absorb starlight.

Recently developed opaque metallic metamaterials are only tens of nanometers thick \citep{Hagglund13}, which approaches theoretical limits \citep{Hagglund10}. At these surface densities (of order 10 g/cm$^3 \times 10$ nm $=10^{-5}$ g/cm$^2$) one Jupiter mass would provide for a shell with $R=9000$ au.  \citet{Lacki16} describes a scheme with dipole antennae that would use even less material for similarly sized Dyson spheres. If such a Dyson sphere could be engineered with significant microwave emissivity (and if the issue of radiation pressure discussed in the next section could be managed), its temperature would be close to $T_{\rm min}$.

This does not mean that the star would be completely hidden. In order to ``blend in'' with the cosmic microwave background at $T=2.7$ K, the Dyson sphere would have to be cooler than $T_{\rm min}$, which would require both actively cooling the Dyson sphere and disposing of the excess stellar luminosity nonthermally. It would also be an incomplete ``cloak'' because the microwave background and foreground are variable across the Dyson sphere's sky, meaning that the Dyson sphere would need to present a different thermal profile to observers in different directions.

\subsection{Other Engineering Issues}

\label{sec:radiation}
Here, we have explored only the outer contours of the problem, probing the physical limits of efficiency and considering only the trade-off of maximum efficiency with Dyson sphere mass. \citet{badescu95} and \citet{Badescu2014} consider the maximum efficiency of an {\it endoreversible} reactor to be a more realistic upper bound. Such a calculation requires additional assumptions about the engineering of the Dyson sphere, and will produce efficiencies below the conservative upper limits here. 

An additional engineering challenge for large Dyson spheres is radiation pressure. We have argued that available mass is an important engineering constraint, but Dyson spheres cannot be made arbitrarily thin. A Dyson sphere enfolding a Sun
like star with a mass surface density near or below 0.8 g/m$^2$ (Equation~(\ref{eq:leaf})) must not only manage the energy and entropy it receives from the star but the momentum as well, lest it be blown away. A Dyson sphere that does dissipative work might pass the momentum on in the form of waste heat (for instance if its outer surface has a much higher effective temperature or microwave emissivity than its inner surface) but thin Dyson spheres doing other forms of work will have to address this in other ways. For scale, one Jupiter mass at this critical surface density would produce a shell with $R= 30$ au and with $T_e = 71$ K.

An additional engineering challenge for massive Dyson spheres is gravity: once the mass of the Dyson sphere begins to approach that of the star, its self-gravity will become dynamically important.

In short, we do not expect Dyson spheres to operate at very low temperatures for a variety of reasons, most importantly the tradeoff between functionality and mass, except perhaps if the only purpose of the Dyson sphere is to be as cold as possible. While the precise optimum is not clear and depends on their function and the cost function of adding material to the Dyson sphere, temperatures of $T_e\gtrsim 100$K are not unreasonable and so are worth searching for, and temperatures near $T_{\rm min}$ would appear to be strongly disfavored on practical grounds, unless their only purpose is to be cold.

\section{Summary}

Dyson spheres are a plausible manifestation of extraterrestrial technology with strong observational consequences. But despite being a well known part of SETI for over 60 years, significant theoretical and observational work remains before upper limits on their existence can be computed.

The idea of a monolithic sphere or other structure, popular in science fiction, is a  canard that does not originate with Dyson. Monolithic spheres are not gravitationally stable, have no obvious utility beyond that provided by a swarm of material, and appear to be mechanically impossible besides.

While the gross observational consequences of Dyson spheres are simple to calculate, the feedback from a Dyson sphere on its star results in detectable changes to the stellar radiation (which we have calculated here under some simplifying assumptions) and stellar structure (which have yet to be calculated.)

Typical sizes and temperatures for Dyson spheres will depend on their function, but the mass required to build a useful Dyson sphere means there is some practical upper limit on their size (and so some lower limit on their temperature). This limit is tightest for Dyson spheres doing purely dissipative work, somewhat lower for Dyson spheres whose primary purpose is to generate low-entropy emission or antimatter, even lower for Dyson spheres doing maximal amounts of computation, and lowest of all for Dyson spheres whose only purpose is to be as large and cool as possible.

This treatment has not considered special cases that may evade some of these conclusions, for instance by considering stars significantly more or less luminous than the sun, exploiting the special thermodynamic properties of black holes, or employing non-electromagnetic physics to use neutrinos or gravitational waves. It has also ignored many practical issues with Dyson spheres, for instance the mechanics of deconstructing planets or stars \citep[e.g.][]{Criswell85}, the effects of other material in the system \citep[e.g.][]{DeBiase08}, and the potential that Dyson spheres might persist long after the species that created them are gone \citep[e.g.][]{Arnold13,Cirkovic19}. It has also largely considered steady-state cases; the dynamics of stellar feedback and the construction or destruction of a Dyson sphere might also have observable consequences. 

Sixty years after Dyson's original suggestion, there remains much work to do on the theory of his eponymous spheres.



\newpage

\acknowledgements

I thank Milan \'Cirkovi\'c and the editors of the Serbian Astronomical Journal for soliciting this review, and for their helpful feedback. I thank Mashall Perrin, Geoff Marcy, David Kipping, Zaza Osmanov, Steinn Sigurdsson, Ibrahim Semiz, Gerry Sussman, and Viorel Badescu for helpful discussions that improved this review. I thank John Gizis for pointing me to Harrop et al.\ and references therein.  

The Center for Exoplanets and Habitable Worlds is supported by the Pennsylvania State University, the Eberly College of Science, and the Pennsylvania Space Grant Consortium. 

This research has made use of 
NASA's Astrophysics Data System Bibliographic Services.




\end{document}